\documentclass[iop]{emulateapj}
\usepackage[varg]{txfonts}
\usepackage{amsfonts}
\usepackage{graphicx}
\usepackage{aas_macros}
\usepackage[colorlinks,urlcolor=blue,citecolor=blue,linkcolor=blue]{hyperref} 
\usepackage{textcomp}
\usepackage{natbib}
\usepackage[usenames, dvipsnames]{color}

\begin{document}

\title{HESS J1943+213: a non-classical high-frequency-peaked BL Lac object}
\shortauthors{S.M. Straal et al.}
\shorttitle{HESS J1943+213, a non-classical HBL object}
\newcounter{aff}
\setcounter{aff}{0}

\author{
S.~M.~Straal \altaffilmark{\ref{uva},\ref{astron}},
K.~\'E.~Gab\'anyi \altaffilmark{\ref{fomi},\ref{konkoly}},
J.~van~Leeuwen  \altaffilmark{\ref{astron},\ref{uva}},
T.~E.~Clarke \altaffilmark{\ref{nrl}},
G.~Dubner \altaffilmark{\ref{argentina}},
S.~Frey \altaffilmark{\ref{fomi}},
E.~Giacani \altaffilmark{\ref{argentina}},
Z.~Paragi \altaffilmark{\ref{jive}}
}

\altaffiltext{1}{\refstepcounter{aff}\label{uva}\refstepcounter{aff}Anton Pannekoek Institute for Astronomy, University of Amsterdam, Science Park 904, PO Box 94249, 1090 GE Amsterdam, The Netherlands; \href{mailto:s.m.straal@uva.nl}{s.m.straal@uva.nl}}
\altaffiltext{\theaff}{\label{astron}\refstepcounter{aff}ASTRON, the Netherlands Institute for Radio Astronomy, PO Box 2, 7990 AA, Dwingeloo, The Netherlands}
\altaffiltext{\theaff}{\label{fomi}\refstepcounter{aff}F\"OMI Satellite Geodetic Observatory, P.O. Box 585, H-1592 Budapest, Hungary}
\altaffiltext{\theaff}{\label{konkoly}\refstepcounter{aff}Konkoly Observatory, MTA Research Centre for Astronomy and Earth Sciences, PO Box 67, H-1525 Budapest, Hungary}
\altaffiltext{\theaff}{\label{nrl}\refstepcounter{aff}U.S. Naval Research Laboratory, 4555 Overlook Avenue SW, Washington, DC 20375, USA}
\altaffiltext{\theaff}{\label{argentina}\refstepcounter{aff}Instituto de Astronom\'{\i}a y F\'{\i}sica del Espacio (IAFE, CONICET-UBA), CC 67, Suc. 28, 1428 Buenos Aires, Argentina}
\altaffiltext{\theaff}{\label{jive}\refstepcounter{aff}Joint Institute for VLBI Eric (JIVE), Postbus 2, 7990 AA, Dwingeloo, The Netherlands}

\setcounter{aff}{0}
\setcounter{footnote}{0}

\begin{abstract}
HESS~J1943+213 is an unidentified TeV source that is likely a high-frequency-peaked BL Lac (HBL) object but also compatible with a pulsar wind nebula (PWN) nature. 
Each of these enormously different astronomical interpretations is supported by some of the observed unusual characteristics. 
In order to finally classify and understand this object we took a three-pronged approach, through time-domain, high angular resolution, and multi-frequency radio studies. 
First, our deep time-domain observations with the Arecibo telescope failed to uncover the putative pulsar powering the proposed PWN. We conclude with $\sim$70$\%$ certainty that HESS J1943+213 does not host a pulsar. 
Second, long-baseline interferometry of the source with 
e-MERLIN at 1.5- and 5- GHz, shows only a core, a point source at $\sim\,$1 -- 100 milli-arcsecond resolution.
Its 2013 flux density is about one-third lower than detected in 2011 observations with similar resolution. This radio variability of the core strengthens the HBL object hypothesis. 
More evidence against the PWN scenario comes, third, from the radio spectrum we compiled. The extended structure  follows a power-law behavior with spectral index $\alpha = -0.54\pm 0.04$ while the core component is flat spectrum ($\alpha = -0.03\pm0.03$). In contrast, the radio synchrotron emission of PWNe predicts a single power-law distribution. Overall we rule out the PWN hypothesis and conclude the source is a BL Lac object. The consistently high fraction (70$\%$) of the flux density from the extended structure then leads us to conclude that HESS J1943+213 must be a non-classical HBL object.

\end{abstract}

\maketitle
\setcounter{footnote}{0}

\section{Introduction}
The unidentified $\gamma$-ray source HESS~J1943+213 is intriguing because of its low Galactic latitude ($-1\fdg29$). It could be the first BL Lac object to be observed through the Galactic plane. Proving on the other hand that HESS~J1943+213 is a pulsar wind nebula (PWN), residing in a supernova remnant (SNR) shell at the outer edge of our Galaxy, could help solve the long-standing problem of the missing Galactic SNRs. With a Galactic supernova rate of one per $30-50$ years \citep{tammann1994}, and a remnant radio-lifetime average of $\geq6\times10^4$ years \citep{frail1994}, there should be $\sim1.6\times10^3$ SNRs at any given time. In reality the detected SNRs make up only $\sim$18$\%$ of this number \citep{green2014}.

HESS~J1943+213 was first discovered as the hard X-ray source \mbox{IGRD~ J19443+2117} by \cite{landi2009} with {\em INTEGRAL}. \cite{landi2009} identified the source in several wavelength bands. In the X- and $\gamma$-rays and in the infrared (IR; J-, H-, and K-band) it corresponds to the source \mbox{2MASS~J1943562+2118233}, and in radio at 1.4\,GHz to \mbox{NVSS~J194356+211826}. A power law was fitted to the {\em INTEGRAL} data in the 0.9\,$-$\,100\,keV energy band, which provides evidence for absorption in excess of the Galactic value and a slope typical to the spectral indices of an active galactic nucleus (AGN). The source is highly absorbed in the IR J, H, and K band, which gives an E\textit{(B$-$V)} also in excess of the Galactic value. The excess of Galactic absorption provides evidence for the source's extragalactic nature and in combination with the X-, $\gamma$-rays and radio data, \citet{landi2009} propose HESS\,J1943+213 is a radio-quiet AGN.
\setcounter{footnote}{0}

In order to localize and determine the nature of the source, \mbox{IGR J19443+2117} was followed up in X-rays with {\em Chandra} \citep{tomsick2009}. {\em Chandra}'s higher positional accuracy confirmed the association between the X-ray, IR and radio sources. An absorbing column density significantly higher than the Galactic value ($\rm{N_{H}/N_{H_2}} = 0.84/0.054$) provided more evidence for the extragalactic source nature.
\setcounter{footnote}{0}
In 2011 the source was discovered to also emit at TeV energies \citep{HESS2011,cerruti2011}, as HESS J1943+213. Its detection at very high energies (VHE) together with its flat radio spectrum shown by \cite{HESS2011} make it plausible that the source is a PWN. Many unresolved VHE Galactic sources turn out to be PWNe\footnote{TeVCat, an online catalog for TeV Astronomy, \url{http://tevcat.uchicago.edu/}}. Although the source has now been observed at multiple energies and wavelengths, its nature remained unclear. 

Two plausible scenarios remain for the nature of this source. 
It is either a high-frequency-peaked BL Lac (HBL) object, evidenced by its TeV emission and soft VHE spectrum \citep{HESS2011}, or it could be a Galactic PWN \citep[e.g.][]{HESS2011, gabanyi2013} which is supported by the lack of variability \citep{shahinyan2015b}. \cite{HESS2011} also proposed that the source could be a gamma-ray binary, however, this scenario was quickly discarded because no massive companion was detected to a distance limit of $\sim$25\,kpc. This would place the potential binary outside our Galaxy, making it 100\,$-$\,1000 times brighter in X-rays than any known gamma-ray binary. In support, the observed 10s of arcseconds to $\sim1$-arcminute-scale radio structure \citep{gabanyi2013} cannot be explained by colliding winds the binary would produce. The lower limit of the distance of 16\,kpc derived by HI absorption favours its extragalactic nature \citep{leahy2012}, but is inconclusive, because \cite{vallee2008} shows that the furthest spiral arm of the Milky Way reaches distances greater than 20\,kpc. However, the soft TeV spectrum, $\Gamma=3.1\pm0.5$, is softer than of all known PWNe \citep{kargaltsev2010} and argues in favour of the blazar hypothesis \citep{HESS2011}. Also, \cite{peter2014} interpret the K-band counterpart to be a massive elliptical galaxy, with only 10$\%$ chance it being a star and found a weak 5.1$\sigma$ detection above 1\,GeV of the counterpart of HESS~J1943+213 in 5 years of {\em Fermi} data. This supports the blazar hypothesis as most of the blazars are detected by {\em Fermi} \citep{piner2014}.
Recent {\em VERITAS} observations, conducted by \cite{shahinyan2015b}, show no brightness variability, however, BL Lac objects are known for their wide range in variability time-scales and intensities. Despite the evidence for its BL Lac nature, the radio spectral index obtained by \cite{HESS2011} is also compatible with the PWN scenario. Supportingly, \cite{piner2014} argue that the brightness of the source in TeV is two orders in magnitude too low to be a HBL object, which would leave the PWN scenario as the only plausible scenario. 
\subsection{PWN hypothesis}
\label{sec:intro_pwn}
The PWN hypothesis is strengthened after the re-analysis of archival VLA large-scale HI data \citep[VGPS;][]{stil2006} in \cite{gabanyi2013}. These data revealed the presence of a shell-like feature of $\sim1$\textdegree diameter, the radio/X-ray/TeV point source near its center\citep[see Fig.~3,][]{gabanyi2013}. This shell-like feature can be interpreted as a consequence a supernova explosion where the central compact source is the PWN, powered by a young pulsar. The expansion suggests the supernova explosion to have occurred $4\,\times\,\rm{10^5}$\,yrs ago. Indeed, if one puts the Crab or 3C58 PWNe at the proposed distance of 17\,kpc \citep{gabanyi2013} they would appear the same size as HESS~J1943+213 seen in the archival VLA data. We then expect the young, energetic pulsar powering the PWN to have a period of 30--300\,ms. At the proposed distance, the dispersion measure (DM) would be of order 500\,pc\,cm$^{-3}$ in this line of sight \citep[from NE2001,][]{cordes2002}.
\subsection{BL Lac hypothesis}
In 2011, \cite{gabanyi2013} performed EVN (European VLBI Network) observations of the source at 1.6\,GHz and found the radio counterpart of the high-energy source at an offset of $3\farcs75$ to the NVSS catalog coordinates. The recovered flux density was 31$\pm$3\,mJy, which is only one-third of the flux density ($95\pm9$\,mJy) recovered simultaneously with the Westerbork Synthesis Radio Telescope (WSRT). This latter corresponds well to the flux density obtained from archival data of the Very Large Array (VLA) taken at 1.4 GHz on 1985 September 30 (project: AH196) and thus shows a discrepancy in flux density between the separate angular scales. Using these observations to compare to, we will discuss the proposed BL Lac nature of the source by imaging the source in order to further investigate its sub-arcsecond radio structure.\\

In this paper we present new time-domain, high-resolution imaging, and continuum  investigations of HESS~J1943+213 (hereafter J1943+213).
In Sect. \ref{sec:observations} we describe the observations and data reduction of all three studies. Section \ref{sec:results} contains our findings. We discuss these results in Sect.~\ref{sec:discussion} before concluding, in Sect.~\ref{sec:conclusion}, on the nature of this intriguing source. 

\section{Observations}
\label{sec:observations}
In order to investigate the nature of the source we took a three-pronged approach. To determine if the source is a PWN, we have performed high-time resolution observations with the Arecibo radio telescope to find the putative pulsar powering the PWN. These will be addressed first. Secondly, we investigate any sub-arcsecond-scale radio structures of the source for which we have obtained e-MERLIN (electronic Multi-Element Remotely Linked Interferometer Network) observations. 
Finally, we discuss flux density measurements obtained by the new survey VLITE (VLA Low Band Ionospheric and Transient Experiment, Clarke et al.\ 2016, in preparation). Together with survey catalog flux density measurements, this low-frequency observation at 340\,MHz will provide information on the continuum spectrum which in turn could help to rule out either the proposed PWN or BL Lac object scenario.

\subsection{Pulsar search}
Arecibo, with its 305-m diameter, is the largest single-dish telescope in the world. Combined with its broadband observing capabilities it provides instantaneous sensitivity and is highly suitable for pulsar searches. The source J1943+213 was observed in two frequency bands to improve the chance of detecting the putative pulsar, given the various frequency-dependent effects. If the source is a PWN in a $\sim1$\textdegree\, SNR shell, it is expected to be at a distance of $\sim$17\,kpc \citep{gabanyi2013}. The free electrons in the line of sight create a dispersive time delay in the putative pulsar signal which is proportional to $f^{-2}$, where $f$ is the observing frequency. This delay is expressed through the dispersion measure (DM), the density of free electrons integrated over distance in the line of sight, given in $\rm{pc\,cm^{-3}}$.
Observing at higher frequencies allows for less dispersive time delay, which increases the signal-to-noise (S/N) ratio of a pulsar signal. 
Beyond this dispersion, the signal will suffer from scattering which is proportional to $f^{-4}$. Scattering stretches the pulse shape, also resulting in a reduction of the S/N ratio. These two effects make observing at low frequencies a challenge.

In contrast, pulsars are known to have a steep power spectrum towards low frequencies, mostly peaking in the range of 200$-$400\,MHz. As the pulsar signal is stronger towards lower frequencies, observing at lower frequencies is a boon. It might enable the detection of the pulsar where at higher frequencies it would be too weak. However, taking the above mentioned effects into account, observing below 1.4\,GHz might smear out most of the signal for a far away, high-DM pulsar.

\subsubsection{Observations with Arecibo}
Arecibo was pointed to the radio counterpart of HESS~J1943+213, \mbox{NVSS J194356+211826} \citep{landi2009} on 2012 June 1. The source was observed in the L-wide band around 1.44\,GHz and in the S-wide band around 2.85\,GHz with the Arecibo Mock spectrometers in single-pixel mode. The field-of-view (FoV) for the pointing in the L-wide band was $3\farcm1\,\times\,3\farcm5$, and in the S-wide band $1\farcm8\,\times\,2\farcm0$. In the L-wide band we observed for 54 minutes with a large total bandwidth of 688\,MHz. This is more than twice as large as the PALFA Survey bandwidth of 300\,MHz \citep{knispel2011}. The PALFA Survey is the Arecibo L-band Feed Array 1.4\,GHz Survey for radio Pulsars, which is currently the most sensitive survey of the Galactic plane.  
In the S-wide band we observed using a slightly smaller bandwidth of 494\,MHz for 69 minutes. More observing details can be found in Table~\ref{table:obsspecs_arecibo}. 

\begin{table}
\begin{center}
\caption{Observing specifications pulsar search with Arecibo\label{table:obsspecs_arecibo}}
\begin{tabular}{l r r}
\tableline\tableline
                                           & L-wide & S-wide\\
\tableline
Central frequency (MHz)       & 1444.1 & 2852.3\\
Total Bandwidth (MHz)         & 688      & 493.6\\ 
Number of Channels             & 4096    & 2940\\
Sample time ($\rm{\mu}$s) & 65.45   & 65.48\\
Total time (s)                      & 3237.6 & 4181.9\\ 
\tableline
\end{tabular}
\end{center}
\end{table}

\subsubsection{Data reduction}
The obtained data were streamed to the Cartesius super computer at SURFsara\footnote{\url{https://www.surfsara.nl/nl/systems/cartesius}}. There it was converted from 16-bit to 8-bit data, and the Mock subbands were combined. The resulting PSRFITS files were further analysed with the pulsar search software PRESTO \citep{ransom2001}. Radio frequency interference (RFI) was masked out of the data. The L-wide data were searched for periodic signals in the DM range from 0 to 1000 $\rm{pc\,cm ^{-3}}$, which is twice the expected DM value (see Sect. \ref{sec:intro_pwn}). The DM range was searched starting with steps of 0.05 $\rm{pc\,cm^{-3}}$ and no down-sampling, up to steps of 0.30 $\rm{pc\,cm^{-3}}$ and a down-sampling factor of 4. The candidate signals were sorted on their S/N ratio and searched down to a detection limit of 4.1$\sigma$, equivalent to $\chi^2 \sim$ 1.90. These detection limits, as output by PRESTO, signify how much the candidate signal deviates from a straight line. 
The same DM range was searched in S-wide with steps of 1.00 $\rm{pc\,cm^{-3}}$ and no down-sampling. Here we could afford to search using larger DM steps, because the frequency dependent smearing is less at higher frequencies. Further analysis was performed following the same steps, where in S-wide all candidate signals were inspected.

For every DM step in both bands, we also looked for single pulses of widths between 0.064 and 10\,ms, down to a S/N = 8.

\subsection{e-MERLIN radio imaging and data reduction}
Radio imaging data were obtained by e-MERLIN, which is a UK-based long baseline interferometry array using six 25-m dishes with the optional inclusion of the Lovell telescope (76\,m diameter). The total network can reach a resolution of 150 milli-arcseconds (mas) in the L-band (1.5\,GHz).

The e-MERLIN observations were obtained at 1.5- and 5-\,GHz \citep[project CY1017;][]{gabanyi2015}. The 1.5-\,GHz observation took place on 2013 December 7 and lasted for 12 hours. From the total observing time, approximately 6\,h was on-source time. Unfortunately, the self-calibration did not provide meaningful solutions in the first few hours of the observation and only $\sim$ 4.3\,h could be used for the imaging of the source. The phase calibrator was J1946+2300. The 5-GHz observation had to be split in three 4.5-h long runs, because due to maintenance, the Jodrell Bank MkII telescope could only participate in observations during the night in October 2013. The three runs were carried out between 2013 October 11 and 14. To cover the missing hour angles, two additional runs were carried out on 2014 June 12 and 13. Except for the last run, when MkII was not involved in the observation, all e-MERLIN telescopes participated. The on-source time was approximately 13-h. The phase-reference calibrator was J1925+2106.

Data reduction was done using the National Radio Astronomy Observatory (NRAO) Astronomical Image Processing System \citep[AIPS,][]{greisen2003}, following the e-MERLIN cookbook version 2.4a.

\subsection{VLITE data reduction}
The National Radio Astronomy Observatory's Very Large Array (VLA) is a 27-antenna interferometer operating between 56\,MHz and 50\,GHz \citep{perley2011}. We obtained data from a new commensal observing system on the VLA called the VLA Low Band Ionospheric and Transient Experiment (VLITE). This system records data for a 64\,MHz bandwidth of the low-band receiver \citep{clarke2011} centered at 352\,MHz for 10 VLA antennas (Clarke et al.\ 2016, in preparation). The correlator is
a custom DiFX correlator \citep{deller2011} operating in real time on the VLITE data stream. Science operations began in November 2014 and VLITE data have been recorded for nearly all pointed VLA observations with primary science programs above 1\,GHz since that time. We searched the VLITE archive and found observations with
J1943+213 within the field of view for two separate observations on 2014 December 4. The observations were split across two phase centers for a combined observing time of 11.1 minutes. 

The VLITE data were calibrated following standard reduction procedures with each observation processed separately before combining the final images. RFI was excised using automated routines. Additional fixed flags were applied to remove known bright RFI and aliasing in the $360-384$\,MHz portion of the spectrum. Next the data were corrected for delay offsets followed by an initial round of calibration. Additional flagging was undertaken following that calibration and a second round of calibration was applied. For our target source, the flux density calibration used 3C286 and 3C48 to set the scale. We have converted measured flux densities to the scale of the other measurements using the known scaling for the flux density calibrators.

Following calibration we attempted to undertake phase self-calibration of the data but found too little flux density in the field to improve the phase solution of the data. Each of the individual pointings was convolved to a matched circular beam (52\arcsec) and then corrected for primary beam attenuation using a recently derived VLITE primary beam
appropriate for the VLA subreflector position of the observations. The two pointings were then combined in the image plane for a final image which has an rms of 36.2\,mJy\,beam$^{-1}$.


\section{Results}
\label{sec:results}
\subsection{Pulsar search}
\begin{figure}
  \includegraphics[width=0.5\textwidth]{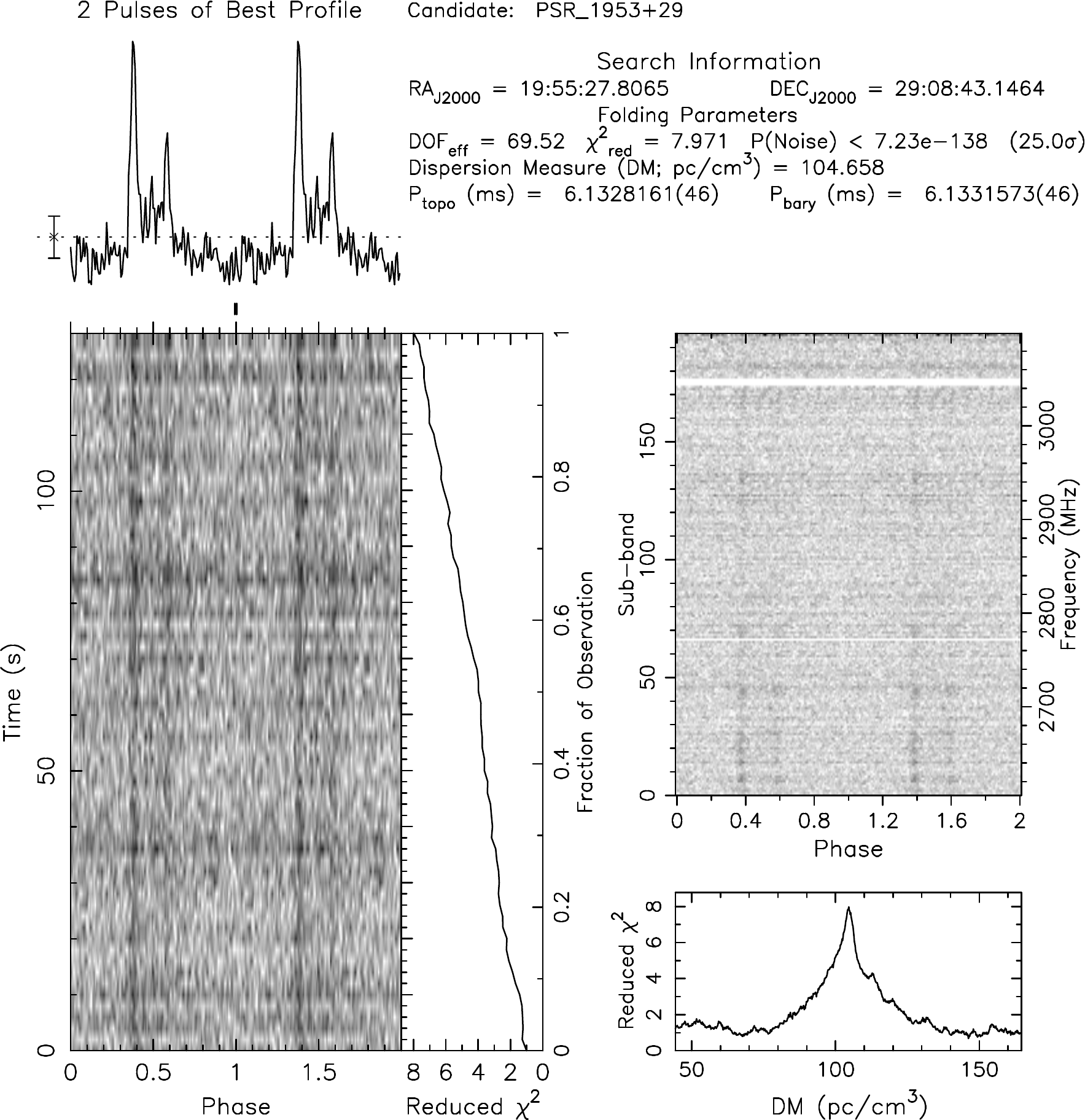}
 \caption{This figure shows the detected pulse signal of the test pulsar B1953+29 observed in the S-wide band as output by PRESTO. The left-hand side shows the cumulative pulse signal and its progression in time. The right-hand side shows (upper) the pulse signal as a function of frequency where the white stripe is masked-out radio frequency interference (RFI). The right-bottom side displays the strength of the pulsar signal as a function of dispersion measure ($\rm{pc \,cm^{-3}}$).\label{fig:testpsr}}
\end{figure}
The  Arecibo  instrument and software setup was verified by observing three known, bright pulsars (PSRs B1937+21, B1953+29, B2016+28) in nearby directions in the sky. All were easily detected (cf.~Fig.~\ref{fig:testpsr}).

Our sensitive search of J1943+213 resulted in $\sim$2000 candidate signals in L-wide and $\sim$1500 candidate signals in S-wide. All these candidate signals were inspected by eye, where we focused on a clean pulse profile as seen in the top left of Fig.~\ref{fig:testpsr} and a peak in the detected DM, as seen on the bottom right of the same figure. A peak in the DM indicates that the signal originates from a specific location. The bottom left graph shows the intensity of the pulse profile as a function of phase and time, which for a single pulsar should look similar to that of the shown test pulsar. No candidate signal down to a detection limit of $\chi^2$=1.90 and 4.1\,$\sigma$ appeared to be a pulsar. For an indication of the thoroughness of these limits, we compare to the PALFA survey. There, the ten most recent detections were on average $\chi^2$ = 8, and 23\,$\sigma$. More specifically the five weakest pulsars in our expected period regime (30\,ms $\le$ P $\le$ 300\,ms) have $\chi^2$= 2.12 -- 4.73 and $\sigma$= 6.1 -- 14.7. These would all have been handily detected. 

To determine the corresponding flux density limits, we use the modified radiometer equation \citep[Eq. \ref{eq:radiometer}, after ][]{dewey1985}:

\begin{equation}
\label{eq:radiometer}
S_{\rm{min}} = \frac{\rm{S/N}}{\rm{Gain}} \frac{T_{\rm{sys}}} {\sqrt{n_{\rm{pol}}\times \tau \times \rm{BW}}}\sqrt{\frac{W}{P-W}}
\end{equation}

In our observations we searched down to S/N = 6 in L-wide and S/N = 4 in S-wide. The gain of Arecibo is 10.5\,K\,Jy$^{-1}$ for L-wide and slightly less (9.5\,K\,Jy$^{-1}$) for S-wide. The frequency-dependent system temperature ($T_{\rm{sys}}$) is 25\,K in L-wide and 32\,K in S-wide. In both bands we recorded both polarization directions ($n_{\rm{pol}}=2$). The bandwidth (BW), expressed in Hz, and observing time ($\tau$) in seconds, are given in Table~\ref{table:obsspecs_arecibo}. $W$ and $P$ are the pulse width and rotation period of the pulsar, respectively. Together, ($W/P$) provide the duty cycle of the pulsar signal (i.e. the 'on'-time of a pulsar). Because W and P are unknown when searching for a new pulsar, we use the average $W_{\rm{50}}/P=0.071$ duty cycle over all PWN pulsars in the ATNF pulsar database \citep{manchester2005}.
We find we were sensitive to signals of 1.90\,$\mu$Jy in L-wide and of 1.45\,$\mu$Jy in S-wide. The pseudo luminosity, $L=Sd^2$, that accounts for the distance, is 0.55\,mJy\,kpc$^2$ (L-wide) and 0.42\,mJy\,kpc$^2$ (S-wide) for the assumed  distance of $d$=17\,kpc. This is 3$-\sim$300 times more sensitive than the luminosity of 88$\%$ of all known pulsars in a PWN.

Next to the search for periodic pulsar signals described above, we have also conducted a single-pulse search. Such a search is sensitive to pulsars that emit only irregularly. 
We identified all 3000 single pulses with S/N ratios above 8, in both bands, and evaluated how their DM versus time signature compared to the shape expected for a pulsar (D. Michilli, \emph{priv.~comm.}). For the best 30 candidates we looked for the dispersed pulse curve in the high time-resolution dynamic spectrum, which would be characteristic of a pulsar. We did not find any such single pulses.

\subsection{e-MERLIN imaging}
\label{sec:results_eMERLIN}
\begin{figure*}
  \includegraphics[width=\textwidth]{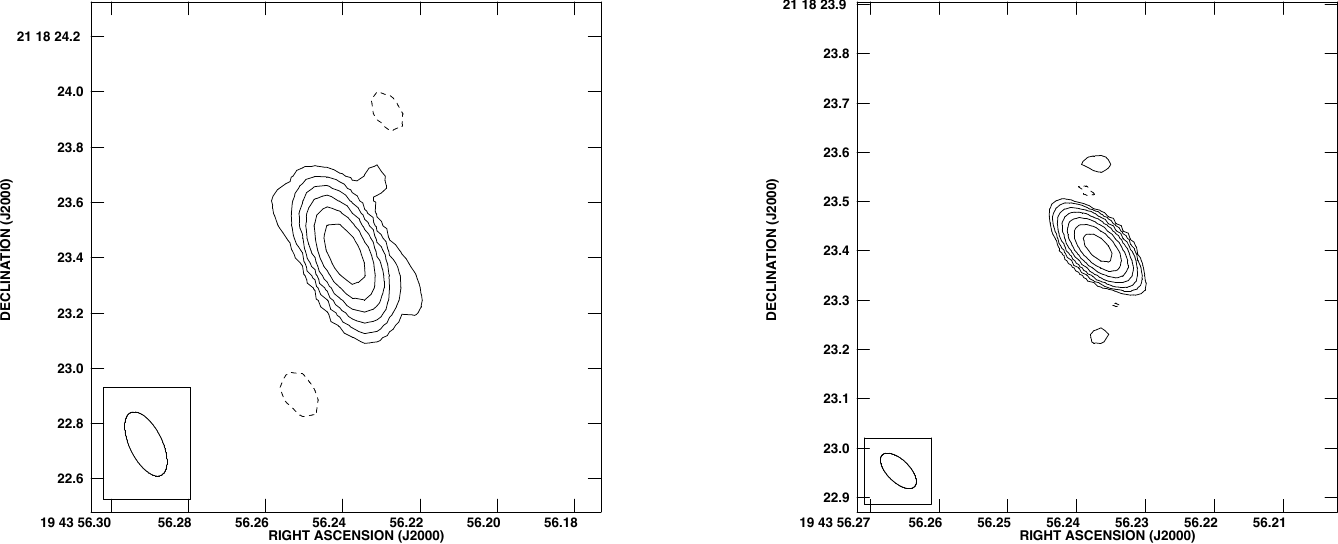}
 \caption{e-MERLIN images of HESSJ943+213. Left-hand side: L-band image. The peak is 19.2\,mJy\,beam$^{-1}$, the beam size is 252\,mas\,$\times$\,117\,mas at a position angle of 26\textdegree. The lowest positive contour is at 0.34\,mJy/beam (7$\sigma$-level), further contours increase with a factor of two. Right-hand side: C-band image. The peak is 19.4\,mJy\,beam$^{-1}$, the beam size is 92\,mas × 45\,mas at a position angle of 46\textdegree. The beams are shown in the lower left corner in each plot. The lowest positive contour is at 0.2\,mJy\,beam$^{-1}$ (7$\sigma$ -level), further contours increase with a factor of two. The dashed contours in both plots indicate the 7$\sigma$ negative contours. \label{fig:e-merlin}}
\end{figure*}

The e-MERLIN data reduction resulted in an unresolved point source at both frequencies, as shown in Fig.~\ref{fig:e-merlin}. None show any large-scale feature visible down to a 7$\sigma$ level (0.34\,mJy\,beam$^{-1}$) in the  $23\arcsec\times 23\arcsec$ L-band FoV. Brightness distribution models were fit to the visibilities with Difmap \citep{shepherd1994} and we found that single, circular, Gaussian components best describe the source at both frequencies. The emission has a flux density of 22.2$\pm$0.7\,mJy at 1.5\,GHz and of 22.4$\pm$0.3\,mJy at 5\,GHz. If we assume that the source did not show variability between the observations taken at the two frequencies, J1943+213 has a flat spectrum. Since the 5-GHz observations were split into several chunks, we were able to compare the source flux density in the different observing runs to check for flux density variability. Specifically we did self-calibration and imaging with using the first 13.5 hours, which were observed on three consecutive days in 2013 October. Separate self-calibration and imaging were done with using only the observations performed in 2014 June. We did not detect significant variability between these two epochs. 

On the other hand, in the L-band J1943+213 was significantly fainter during the e-MERLIN observation in 2013 December compared to the EVN L-band observation in 2011 May \citep{gabanyi2013}. 

\subsection{Radio continuum spectrum of J1943+213}
The radio continuum spectrum of J1943+213 is obtained by combining several radio survey flux density measurements of the source and measurements from observations pointed to the NVSS source, which is the accepted radio counterpart of J1943+213.

\subsubsection{Flux density measurements from radio surveys}
The VLITE observations detect a source at the NVSS position with almost 8-$\rm{\sigma}$ significance. The flux densities at 340\,MHz are 0.23$\pm$0.5\,Jy and 0.36$\pm$0.07\,Jy integrated. From the fit, the structure appears slightly extended (56\arcsec) compared to the 52\arcsec$\times$52\arcsec resolution. The reported extension is only marginally larger than the beam size, thus it is unclear if the source is resolved, and we therefore have chosen to use the \emph{peak} flux density which is appropriate for an unresolved source.

The source is also detected by the re-reduction of the VLA low-frequency Sky Survey \citep[VLSSr,][]{lane2014} at 73.8\,MHz. The source however did not end up in the VLSSr catalog due to a 5-$\rm{\sigma}$ cutoff. Careful data reduction shows that the source is detected with a detection significance just below 4\,$\rm{\sigma}$. The smaller-than-5-$\rm{\sigma}$ detection is strengthened by the spatial match (within 23\arcsec) to the NVSS source. The source is unresolved by the $75\arcsec\,\times\,75\arcsec$ beam of VLSSr. The 73.8-MHz flux density obtained is $0.37\pm0.09$\,Jy where we have corrected for clean bias and included a 12$\%$ flux density scale uncertainty.

Finally, the radio counterpart of J1943+213 is also detected at higher radio frequencies. We find a detection of J1943+213 in the Arcminute Microkelvin Imager Galactic Plane Survey \citep[AMIGPS,][]{perrott2015} at 15.7\,GHz, well within the $\sim$\,5\arcsec error circle of AMIGPS. With its 3\arcmin resolution, the source is detected as a point source with a flux density of $23.5\pm3$\,mJy.

\subsubsection{Radio continuum spectrum}
\begin{figure*}
  \includegraphics[width=\textwidth]{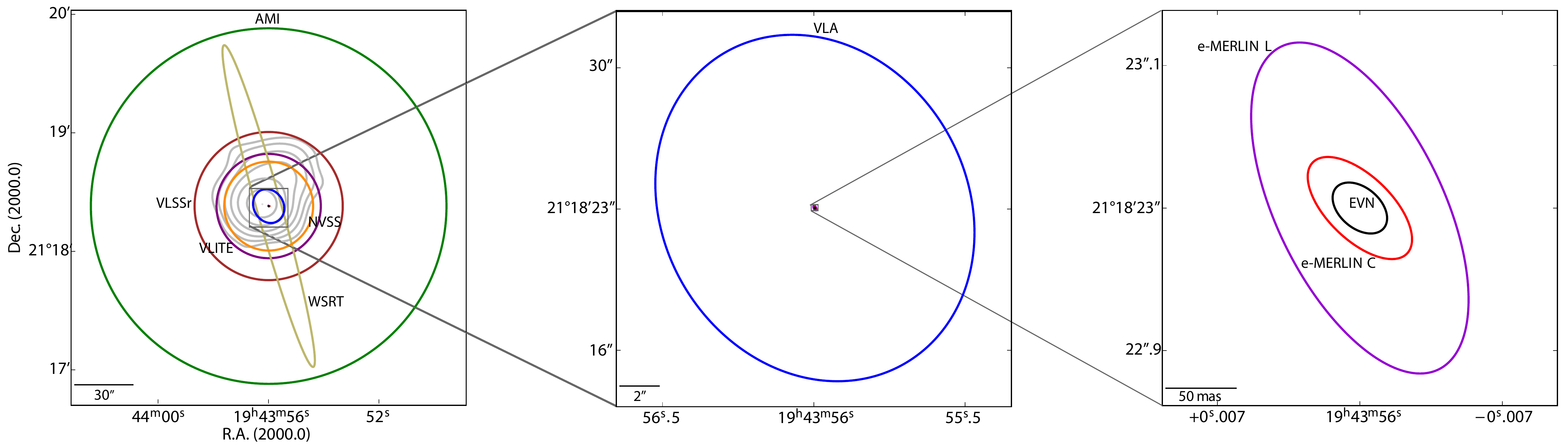}
 \caption{This image shows the FWHM of the beams of the observations of HESS J1943+213 with different radio telescopes or interferometers. The corresponding sizes and observed flux densities can be found in Table \ref{table:fluxes}. In the left panel the VLA 1\farcm1\,$\times$\,0\farcm8 structure \citep{gabanyi2013} is overplotted. The instruments corresponding to the right-most panel are unable to probe the extended structure observed by VLA. \label{fig:radiobeams} }
\end{figure*}

Our radio flux densities, obtained from radio surveys and pointed observations, are listed in Table \ref{table:fluxes}. The flux density measurements have been brought to the absolute flux density scale using the VLA formula with 1999.2 coefficients, and the WSRT flux density measurements using Perley-Butler time dependent coefficients. These are sorted by observing band and thereafter by observing epoch to facilitate the identification of variability in time. To also distinguish between various spatial scales, we list the beam sizes (full width at half-maximum, FWHM). These beam sizes are next shown in Fig.~\ref{fig:radiobeams}, centered on the NVSS catalog coordinates of J1943+213. For the sake of clarity the large, highly elongated Nanc\c{}ay Radio Telescope (NRT) beam \citep{HESS2011} is omitted. Also overplotted is the $1\farcm1\,\times\,0\farcm8$ extended radio structure detected by the VLA at 1.4\,GHz \citep{gabanyi2013}. Figure~\ref{fig:radiobeams} shows which observations can be expected to resolve the extended radio structure. Both AMIGPS and VLSSr see J1943+213 as a point source. Given its match to the structure size, the VLITE beam may or may not marginally resolve the source.

\begin{figure}
  \includegraphics[width=0.5\textwidth]{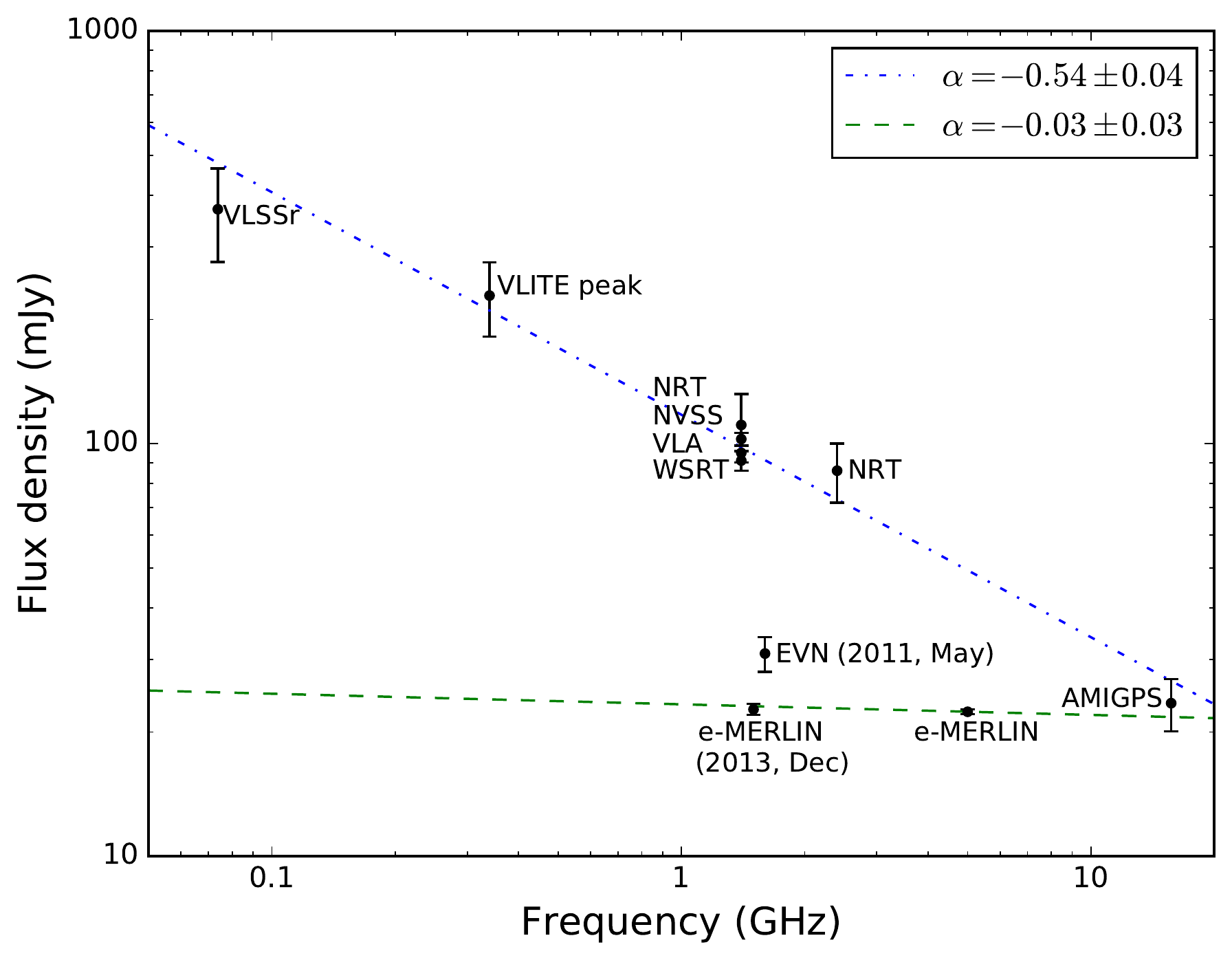}
 \caption{Radio flux densities obtained from surveys or pointed observations plotted against their observed frequency. The observations of VLSSr, VLITE, NRT, NVSS, VLA, WSRT, and AMIGPS probe the sum of the extended structure and core component. The e-MERLIN and EVN observations only probe the core component (see Fig.~\ref{fig:radiobeams}). The epochs for the L-band EVN and e-MERLIN observations are shown, to indicate the variability between both observations. The blue dash-dotted line is the power-law fit to the flux densities of the extended structure and core component. The green, dashed line is the power-law fit to the core component flux density measurements. \label{fig:powerlaw}}
\end{figure}

The observations of J1943+213 can be sorted in two groups based on the structure size each observation was able to probe. Eight observations are able to see the extended structure of J1943+213 (shown in the two left panels of Fig. \ref{fig:radiobeams}). The other group (EVN and e-MERLIN) is only able to observe the mas-scale structures which we hereafter call the core. The observations were unable to measure small-scale structures larger than 2\arcsec, limited by the maximum angular scale that e-MERLIN can recover in the L-band. The right panel of Fig.~\ref{fig:radiobeams} is a magnification of the left panel by a factor of approximately 700. The EVN and e-MERLIN observations were unable to probe the observed VLA structure.
In Fig.~\ref{fig:powerlaw} we plot the obtained flux densities against their observed frequency. We then fit power laws ($f_{\rm{x}} = bx^{\alpha}$, where $b$ is the offset and $\alpha$ the index) to each of the two above mentioned groups. Observations sensitive to the sum of the extended structure plus the core follow a power law with index $\alpha = -0.54\pm 0.04$. Although the small-scale structure appears to be variable in time (see Section \ref{sec:results_eMERLIN}), the flux density variability falls well within the error bars of the larger-scale flux density measurements and therefore we do not expect to be able to detect variability in the large-scale structure.
The smaller-scale observations that resolved out the extended structure show a flat radio spectrum. Although we observed variability between the 2011 EVN observation and the 2013 e-MERLIN observation, we fit a power law to all observing epochs in order to obtain the average spectral index $\alpha = -0.03\pm0.03$. 
As visible in Fig.~\ref{fig:powerlaw}, this flat core spectrum completely accounts for the AMIGPS flux density; there is no 16-GHz extended-structure emission. 

\begin{table*}
\begin{center}
\caption{Obtained flux densities from different radio instruments\label{table:fluxes}.}
\begin{tabular}{l r c c c c}
\tableline\tableline
Instrument & Band & Flux density & Beam size &  Observation date & Ref.\\
                   &         & (mJy)&             &                  &   \\
\tableline
VLSSr & 73.8\,MHz & $370\pm94$&$75\arcsec$ &  2003 September 20 & This work\\
VLITE & 340\,MHz & $229\pm47$ & $52\arcsec$ & 2014 December 04 & This work\\
VLA & 1.4\,GHz & $91\pm5$ & $17\farcs8\times15\farcs1$ &  1985 September 30& a\\
NVSS & 1.4\,GHz & $102.6\pm3.6$ & $45\arcsec$ & 1993$-$1996 & b\\
NRT & 1.4\,GHz & $111\pm20$ & $2\farcm94\,\times\,20\farcm6$ &  2010 March$-$May  & c\\ 
EVN & 1.6\,GHz & $31\pm3$ & 43.9\,mas\,$\times$\,28.5\,mas &  2011 May 18 & a\\
WSRT & 1.6\,GHz & $95\pm9$ & $2\farcm82\,\times\,0\farcm21$ &  2011 May 18 & a\\
e-MERLIN & 1.5\,GHz & $22.7\pm0.7$ & 117\,mas\,$\times$\,252\,mas &  2013 December 7 & This work\\
NRT & 2.4\,GHz  & $86\pm14$  &$1\farcm82\,\times\,15\farcm6$ &  2010 March$-$May & c\\
e-MERLIN & 5\,GHz & $22.4\pm0.3$ & 45\,mas\,$\times$\,92\,mas &  2013 October $\&$ 2014 June & This work\\
AMIGPS & 15.7\,GHz & $23.5\pm3.4$ & $3\arcmin$  &  2011 April 12$-$17 & d\\
\tableline
\end{tabular}
\tablecomments{The flux density measurements have been brought up to the absolute flux density scale using the VLA formula with 1999.2 coefficients, and the WSRT flux density measurements using Perley-Butler time dependent coefficients.}
\tablerefs{(a) \cite{gabanyi2013}, (b) \cite{condon1998}, (c) \cite{HESS2011}, (d) \cite{perrott2015}}
\end{center}
\end{table*}

\section{Discussion}
\label{sec:discussion}
\subsection{PWN hypothesis}
\label{sec:discussion_pwn}
Our Arecibo observations would be able to detect 88$\%$ of all known pulsars in a PWN, placed at the assumed distance of $\sim$17\,kpc for J1943+213, and if beamed towards us. \cite{ravi2010} show that the radio beaming fraction of $\gamma$-ray pulsars is close to unity for the highest energetic pulsars and goes down to $\sim$0.5 for lower-energetic $\gamma$-ray pulsars. From the ATNF pulsar catalog \citep{manchester2005} we obtain that three-quarters of the pulsars in a PWN emit at high energies (i.e. X-rays and $\gamma$-rays). They are also amongst the highest energetic pulsars and are thus expected to have a large beaming fraction. This means they are observable over a large range of lines of sight. Using the observationally deduced beaming fraction average of $0.75\pm0.25$ and our obtained sensitivity, we can conclude with $0.7\pm0.2$ certainty that there is no pulsar in this source and it is therefore not a PWN. In support, we find that the radio continuum emission can be best described by two components, whereas the radio synchrotron emission of PWNe is described by a single power-law distribution with typical indices between $-0.3$ and 0 \citep{gaensler2006}.

In principle, a two-component radio continuum emission could also be explained by a \emph{composite type} SNR, where PWN is immersed in a faint SNR shell. This is seen in e.g. the composite SNR G292.0+1.8 \citep{gaensler2003}, where the contrast in surface brightness between the core and the surrounding plateau is one order of magnitude, and the respective spectral indices are $\alpha = -0.05$ and $\alpha = -0.5$. There is, however, a discrepancy in scale: if one were to observe G292.0+1.8 and similar composite SNRs (G0.9+0.1 \citet{dubner2008}; HESS~J1818-154 \citet{HESSJ18182014}; or LMC~0540-69.3 \citet{brantseg2014}) at the assumed distance to J1943+213 of 17\,kpc, they would be a factor of $\sim 10^3$ larger than observed for J1943+213. Also, in the known sources, the SNR/PWN size ratio is 2.2 -- 7.5, whereas for J1943+213 this ratio would be approximately $4\times10^3$. This would imply that the expansion of the SNR is orders of magnitude larger than that of the PWN. Overall it appears highly unlikely that J1943+213 is a SNR of the composite type.

\subsection{Blazar hypothesis}
\label{sec:discussion_blazar}
The compactness of the source and the observed radio flux density variability agree with the proposed BL Lac nature of the source. Observations of J1943+213 taken 2014 November with EVN at
1.6\,GHz reveal a complex core-jet morphology (Akiyama et al. 2016). The core brightness-temperature lower limit of $1.8 \times 10^9$ K is consistent with a BL Lac. Akiyama et al. (2016) recover 42 mJy of flux density, 10 mJy higher than found in our EVN observation 2011, confirming the source variability. Compared to the low-resolution observations significant amounts of flux density remain missing.
The large-scale emission observed by the VLA in 1985 and again confirmed by the WSRT observation during our EVN run in 2011, seems to be resolved out in the EVN and e-MERLIN observations. Compared to the WSRT, archival VLA, NRT and NVSS values, approximately 70\,mJy flux density is missing at L-band. This cannot be explained by source-intrinsic changes, since the WSRT measurement was obtained as part of our EVN observation and thus has to come from the large-scale emission. 

BL Lac objects are a sub-class of blazars in which the spectral energy distribution (SED) is dominated by the emission from a relativistic jet pointing close to our line of sight \citep{urry1995}. In the radio, we witness a highly variable emission, compact on mas scales just like in flat-spectrum quasars. In the optical regime, however, a striking difference is the lack of broad emission lines. This is believed to be the result of the different nature of the accretion flows: quasars having a geometrically thin but optically thick accretion disc and accrete close to critical Eddington rates, while BL Lacs accrete at a much lower rate through a thick accretion disc and are radiatively inefficient \citep{maraschi2003,ghisellini2008}. Therefore, instead of the classical distinction between the two classes based on the equivalent width of their emission lines, a physically more motivated approach is the division by the Eddington luminosity of the broad-line regions with a division line $L_{\rm{BLR}}/L_{\rm{Edd}}< 5\times10^{-4} $\citep{ghisellini2011}. One may expect then that at the lower end of BL Lac accretion rates, the SED of the galaxy will not necessarily be dominated by the relativistic jet from the active nucleus, and the radio emission from the Doppler-boosted jet base, optically thick to synchrotron emission in the radio (resulting the flat spectrum and mas-scale compact structure) may not necessarily dominate over the large-scale radio emission. Indeed, \citet{giroletti2004, giroletti2006} have shown for low-redshift ($z<$0.2) BL Lac objects selected from flux density limited samples in the radio, have weak radio cores and a variety of radio morphologies on kpc scales (jets, halos, secondary compact components). The properties of local BL Lacs are found to be in fact similar to their Fanaroff--Riley type I \citep{fanaroff1974} radio galaxy parent population. Further to this, a sample of 42 low-redshift ($z<$0.2) BL Lac objects, selected based on their broad-band properties (with no constraints on their radio and gamma-ray emission), revealed a number of $\lq\lq$non-classical" BL Lacs that have low source compactness, core dominance, and/or show no $\gamma$-ray emission and have steep radio spectra \citep{liuzzo2013}.
From combining observations of surveys and pointed observations we find that the radio continuum emission of the extended structure follows a steep power law with spectral index $\alpha = -0.54\pm 0.04$. At $\sim$16\,GHz the contribution of the extended structure to the total emission becomes almost negligible as can be seen in Fig.~\ref{fig:powerlaw}. The measured flux density at $\sim$16\,GHz agrees well with the expected flux density for the flat-spectrum ($\alpha = -0.03\pm0.03$) core. Our obtained index for the extended structure agrees well with the index ($\alpha_R = -0.59\pm0.16$) that the \cite{HESS2011} obtained from their NRT observations, but is somewhat steeper than the obtained index of $\alpha_R = -0.39$ from archival observations of the source at positions that differ 1\farcs7$-$40\arcsec of the NVSS catalog coordinates. This can be attributed to the fact that we only take measurements into account from which we can confirm the spatial match to the NVSS source, which might not be the case for some of the nine archival observations used by \citet{HESS2011}.
The high-energy spectrum of J1943+213 also corresponds well with the HBL object hypothesis. \citet{peter2014} computed a fit for the spectral energy distribution from radio up to TeV energies. The spectrum can be well described by a self-synchroton model with a black-body component for the host galaxy \citep[see Fig.~7 in][]{peter2014}. The radio fluxes listed in Table \ref{table:fluxes} correspond well to the model provided by \cite{peter2014}. When limiting only to the flux densities attributed to the core structure (typically the dominant emission component in HBL objects), we find that these even agree better with the fit than the single NVSS radio point \cite{peter2014} used.

One may conclude that the compact radio emission in HESS J1943+213 might represent another $\lq\lq$non-classical" BL Lac related low-power AGN activity in a low-redshift galaxy. We note however that this would still be an extreme case, with the lowest core dominance ever witnessed in a BL Lac object (only $\sim30\%$ flux density in the core). Other differences are that the extent of the extended radio emission is larger ($\sim$ 1\arcmin; but the redshift and therefore the linear size is not known), and that there is no apparent connection found yet between the mas-scale core and the diffuse radio emission, in spite of our attempt to reveal it with e-MERLIN that is sensitive to structures between $\sim$100 mas to 2\arcsec.

\citet{peter2014} gave a possible redshift interval for the source as $0.03\leq z\leq 0.45$. Using these values and assuming a standard flat $\Lambda$CDM cosmology ($H_0=67.3\,\rm{km\,s}^{-1}\,\rm{Mpc}^{-1}, \rm{\Omega_{m}=0.315}$, \citealt{planck2014}), the extended 1-arcmin structure seen in the archival VLA image can be translated into a linear size of 38\,kpc $-$ 358\,kpc. Additionally, we can set a lower size limit for the extended feature of 2 arcsec from our L-band e-MERLIN observation. This translates into 1.3\,kpc $-$ 11.9\,kpc. Thus the extended structure of J1943+213 responsible for the $\sim$ 70\,mJy flux density missing from the EVN observation should be between 1.3\,kpc and 36\,kpc in linear size, if the source is close, at a redshift of $z =0.03$, or between 11.9\,kpc and 349\,kpc in linear size, if the source is more distant at a redshift of $z=0.45$. However, since the source is at low Galactic latitude ($-1\fdg29$), it cannot be discarded that there is a chance alignment along the line of sight of a Galactic non-thermal source, our $\lq\lq$large-scale" structure, and the compact presumably extragalactic source, implying that HESS~J1943+213 is a core-dominated BL Lac object without extended structure.

\section{Conclusion}
\label{sec:conclusion} 
In order to classify HESS J1943+213, we have presented imaging and time-domain observations. Our non-detection of a pulsar in the time-domain observations allows us to conclude with $\sim$70$\%$ probability that there is no pulsar in this source. Together with the two components with different power-law radio spectra we obtain from the radio continuum flux densities of the source, which is unlikely to represent an immersed PWN in a SNR shell, and previous arguments against the PWN scenario such as the overly soft X-ray spectrum, we conclude that the source is not a PWN.

The HBL object classification is further strengthened by our new e-MERLIN imaging observations and the radio continuum spectrum we obtain from flux density measurements of surveys (e.g. VLITE and VLSSr) and pointed observations. From the continuum spectrum and our observations with different angular resolutions we find that the object can be best described by two structures: an extended structure with a somewhat steep spectrum ($\alpha=-0.54\pm 0.04$) and a flat-spectrum core with $\alpha=-0.03\pm0.03$. Such a structure is common to BL Lac objects.
Quite striking however is the large 70\% flux density fraction originating from the extended structure, unseen in typical HBL objects. We conclude that HESS J1943+213 is most likely a non-classical high-frequency peaked BL Lac object. Alternatively, since the source is at low Galactic latitude, we cannot rule out that the compact emission is of extragalactic origin while the extended emission is from the Galaxy in the same the line of sight and is physically unrelated to the compact component.

 \section*{Acknowledgements}
We thank Daniele Michilli for providing us the sifting algorithms for the single-pulse search, Adam Deller for  interesting discussions, and Pierre E. Belles from the e-MERLIN staff for extensive support and help in data reduction.
The e-MERLIN is a National Facility operated by the University of Manchester at Jodrell Bank Observatory on behalf of the UK Science and Technology Facilities Council (STFC). 
The Arecibo Observatory is operated by SRI International under a
cooperative agreement with the National Science Foundation
(AST-1100968), and in alliance with Ana G. M\'endez-Universidad
Metropolitana, and the Universities Space Research Association. This work made use of data from the VLA Low-band Ionospheric and Transient Experiment (VLITE). Construction and installation of VLITE was supported by the Naval Research Laboratory (NRL) Sustainment Restoration and Maintenance funding
The research leading to these results has received funding from the European Research Council under the European Union's Seventh Framework Programme (FP/2007-2013) / ERC Grant Agreement n. 617199 (ALERT) and n. 283393 (RadioNet3); from the Netherlands Research School for Astronomy (NOVA4-ARTS); and from the Hungarian Scientific Research Fund (OTKA K104539). Basic research in radio astronomy at the Naval Research Laboratory is funded by 6.1 Base funding. GD and EG are members of the CIC (CONICET, Argentina) and acknowledge support from ANPCyT and CONICET funding.
Part of this work was carried out on the Dutch national e-infrastructure with the support of SURF Cooperative. Computing time was provided by NWO Physical Sciences.

\bibliographystyle{yahapj} 
\bibliography{draft_J1943+213_rev4}

\end{document}